\documentclass[aps,prl,superscriptaddress,reprint,showpacs,amsfonts,amsmath]{revtex4-1}
\usepackage[final]{graphicx}

\usepackage[usenames,dvipsnames]{xcolor}
\usepackage[normalem]{ulem}

\DeclareMathOperator{\sgn}{sgn}

\begin{document}
\title{Qualitative Features at the Glass Crossover}
\author{Tommaso Rizzo}
\affiliation{Dip.\ Fisica, Universit\`a ``Sapienza'', Piazzale A.~Moro 2, I--00185, Rome, Italy}
\affiliation{IPCF-CNR, UOS Rome, Universit\`a ``Sapienza'', Piazzale A.~Moro 2, I-00185, Rome, Italy}
\author{Thomas Voigtmann}
\affiliation{Deutsches Zentrum f\"ur Luft- und Raumfahrt (DLR), 51170 K\"oln, Germany}
\affiliation{Department of Physics, Heinrich-Heine-Universit\"at D\"usseldorf, Universit\"atsstra\ss{}e 1, 40225 D\"usseldorf, Germany}

\begin{abstract}
We discuss some generic features of the dynamics of glass-forming liquids close to the glass transition singularity of the idealized mode-coupling theory (MCT). The analysis is based on a recent model by one of the authors for the intermediate-time dynamics ($\beta$ relaxation), derived by applying dynamical field-theory techniques to the idealized MCT. Combined with the assumption of time-temperature superposition for the slow structural ($\alpha$) relaxation, the model naturally explains three prominent features of the dynamical crossover: the change from a power-law to exponential increase in the structural relaxation time, the replacement of the Stokes-Einstein relation between diffusion and viscosity by a fractional law, and two distinct growth regimes of the thermal susceptibility that has been associated to dynamical heterogeneities.
\end{abstract}

\pacs{64.70.Q-}

\maketitle

To describe how classical liquids arrest kinetically at the glass transition, is still a controversial issue within statistical physics. Starting from the high-temperature liquid state, it is natural to focus on the dramatic slowing down in the dynamics. Here the mode-coupling theory of the glass transition (MCT) is successful \cite{Goetze99,Goetze09}. It predicts the divergence of structural relaxation times at an ideal glass transition temperature $T_c$. Although the singularity is ``avoided'' in real glass formers, it is still signaled by asymptotic, so-called $\beta$-relaxation scaling laws for the dynamics at $T\approx T_c$ \cite{Franosch97}, or a square-root singularity in the scattering intensities \cite{Yang95prl,Yang95a,*Yang95b,Yang96,Adichtchev02,Petzold10}.

On the low-temperature side, the replica method describes the properties of the (metastable) glassy states below $T_c$ \cite{Mezard99,Franz12} yielding fewer but similar quantitative predictions \cite{Szamel10,Rizzo13b}. Random-first-order transition (RFOT) theory \cite{Kirkpatrick89,Bouchaud04} builds on top of replica results 
by advocating entropic nucleation processes to restore ergodicity  below $T_c$ and predicts a debated divergence of a correlation length below the calorimetric glass transition.

Several attempts have been made at ``extended MCT'' to incorporate such physics below $T_c$: introducing further relaxation channels to the MCT equations \cite{Goetze87,Sjoegren90,Fuchs92,Goetze00,Chong08,Chong09,Domschke11}, taking into account higher-order factorization of many-particle density modes \cite{Mayer06,Mayer14}, considering generalized-hydrodynamic arguments \cite{Cates06}, leading to time-dependent coupling coefficients \cite{Greenall07}, using ideas of RFOT theory \cite{Bhattacharyya05}, or within ``na\"i{}ve'' MCT \cite{Schweizer03,*Saltzman03,Schweizer05,Saltzman08}. By and large, all remained rather empirical.

One of us \cite{Rizzo13} has developed a new approach to ideal MCT based on the fact that close to $T_c$, fluctuations will become important. The study of these fluctuations in a full-fledged dynamical context provides a crucial improvement on earlier static treatments \cite{Franz11}. After a  complex field theoretical computation the original problem is mapped onto a rather intuitive model that extends the $\beta$-scaling laws (rather than full MCT, due to saddle-point approximations in its derivation) to a spatially inhomogeneous case where the distance to $T_c$ becomes a spatially fluctuating variable. The corresponding stochastic $\beta$-relaxation (SBR) theory stands out from typical extended MCT, since it was derived \emph{without} the \emph{ad hoc} assumption of ergodicity-restoring processes. It describes the avoidance of the ideal-MCT glass transition as arising from the coexistence of liquid-like and already solid-like regions in a system close to dynamic arrest.

Here we demonstrate that the SBR theory naturally explains three generic features found in virtually all glass-forming systems around $T_c$. The first is a change in the way the structural relaxation time increases, from power-law-like according to MCT above $T_c$, to exponential below $T_c$. The second concerns the decoupling of the single-particle diffusion coefficient $D$ from the collective processes that determine viscosity $\eta$. A Stokes-Einstein (SE) relation, $D\cdot\eta\sim\text{const.}$, happens to be valid in most dense liquids above $T_c$. Close to and below $T_c$ one often finds instead a ``fractional SE'' relation \cite{Chang94,Ediger00}, $D\cdot\eta^{x}\sim\text{const.}$ with $x<1$, that has not been satisfactorily explained yet. In the SBR theory, it emerges as the consequence of the MCT scaling laws close to $T_c$.

The third feature concerns the growth of dynamical susceptibilities. MCT describes the average dynamics in terms of two-point correlation functions. Fluctuations around this average can be related to a dynamical susceptibility whose amplitude diverges as a power law as $T_c$ is approached \cite{Toninelli05}. At $T_c$, a crossover to slower logarithmic growth has been rationalized empirically \cite{DalleFerrier07,Berthier11}. SBR theory explains these two scaling limits.

Around $T_c$ the normalized density correlation function $\Phi(k,t)$ for wave numbers $k$ related to short-wavelength fluctuations remains close to the ideal nonergodic contribution $f(k)$ over a certain time window, and there the difference obeys the leading-order factorization
\begin{equation}
  \Phi(k,t)-f(k)=G_\sigma(t)\xi_c^R(k)\,.
\end{equation}
$\xi_c^R(k)$ is a critical amplitude that depends only on wave vector. The dynamical scaling function $G_\sigma(t)$ is obtained within SBR as the average over a slowly varying real field in space-time related to long-wavelength fluctuations,
\begin{equation}
  G_\sigma(t)=\left[\phi_{\sigma+s}(x,t)\right]\,.
\end{equation}
Square brackets indicate an average over the fluctuating scalar field $s(x)$, which has a Gaussian distribution with zero mean and variance $\Delta\sigma^2$:
$[s(x)]=0$ and
$[s(x)s(y)]=\Delta\sigma^2\,\delta(x-y)$.
The time-scale-invariant equation
\begin{multline}\label{phixt}
  \sigma+s(x)=-\alpha\nabla^2\phi(x,t)+(1-\lambda)\phi^2(x,t)
  \\
  + \int_0^t \left(\phi(x,t-t')-\phi(x,t)\right)\frac{d}{dt'}\phi(x,t')\,dt'
\end{multline}
determines $\phi_{\sigma+s}(x,t)$. The uniform initial
condition $\lim_{t\to0}\phi(x,t)t^a=1$ fixes a unique solution.
Without fluctuations, i.e., for $s(x)\equiv0$, Eq.~\eqref{phixt} recovers the well-known $\beta$-scaling equation of ideal MCT (where the gradient term is irrelevant).
This limit is also recovered for $\alpha\to\infty$, since this also suppresses spatial fluctuations.
Here $\sigma$ is the MCT distance parameter, asymptotically proportional to $(T_c-T)/T_c$, and $\lambda$ is the exponent parameter characterizing the (material-dependent) shape of the scaling function.
We discuss in the following a simplified model, setting $\alpha=0$ but $s(x)\not\equiv0$. It can be argued that this captures the qualitative features of the full Eq.~\eqref{phixt}, while still being amenable to analytical treatment.

Recall some asymptotic statements of ideal MCT.
The solution for $s(x)\equiv0$ is
$G_\sigma(t)=|\sigma|^{1/2}g_{\sgn\sigma}(t/t_\sigma)$, where
$t_\sigma=|\sigma|^{-1/2a}$ is the $\beta$-relaxation time that diverges
upon approaching $T_c$ (where $\sigma\to0$) and $g_\pm(\hat t)$ is the
homogeneous solution of Eq.~\eqref{phixt} with $\sigma=\pm1$.
On this diverging time scale,
$\hat t=t/t_\sigma$,
one has $g_\pm(\hat t)\sim{\hat t}^{-a}$ for $\hat t\to0$,
and long time asymptotes that differ between the glass ($\sigma>0$)
and the liquid ($\sigma<0$): $g_+(\hat t)\sim(1-\lambda)^{-1/2}$
but $g_-(\hat t)\sim-B_\lambda{\hat t}^b$
with a (tabulated) constant $B_\lambda$, for
$\hat t\to\infty$. The exponents $a$ and $b$ are determined by
$\Gamma^2(1-a)/\Gamma(1-2a)=\lambda
=\Gamma^2(1+b)/\Gamma(1+2b)$ (with Euler's Gamma function).
In the scaling function for the liquid, there emerges a second, more strongly diverging time scale, as $G_\sigma(t)\sim-(t/t_\sigma')^b$ with $t_\sigma'=B^{-1/b}|\sigma|^{-\gamma}$ and $\gamma=1/2a+1/2b$. The corresponding power law is referred to as the von~Schweidler law \cite{Goetze09} describing the initial relaxation from the plateau. The $\beta$-scaling law is valid as long as the deviation from the plateau remains small, of $O(|\sigma|^{1/2})$.

For $\alpha=0$, the solution of the simplified SBR (sSBR) theory can be written
in terms of the continuous family of ideal-MCT scaling functions, and depends
on the fluctuations only through the noise-to-signal ratio $\Delta\sigma/\sigma$:
\begin{equation}\label{gsigt}
  G_\sigma(t)%=\left[\phi^0_{\sigma+s}(t)\right]
  =\int_{-\infty}^\infty\frac{ds}{\sqrt{2\pi}\,\Delta\sigma}
  e^{\frac{-(s-\sigma)^2}{2\Delta\sigma^2}}|s|^{1/2}
  g_{\sgn s}(t|s|^{1/2a})\,.
\end{equation}
This solution always describes an ultimate decay from the nonergodicity
plateau as $G_\sigma(t)\to-\infty$ for $t\to\infty$.
Still, a qualitative change occurs around $\sigma=0$. For $\sigma\ll0$ the
long-time asymptote of the integral in Eq.~\eqref{gsigt} is dominated
by ideal-MCT solutions around the center of the Gaussian distribution.
But for $\sigma>0$, the final decay is dominated by contributions from the
tail of the distriution, i.e., by the atypical liquid-like solutions that
occur with exponentially small probability.
Hence, the theory describes the typical signature of an avoided ideal-MCT
glass-transition singularity at $\sigma=0$.

For the numerical calculations
we consider $\lambda=0.75$
(motivated by typical values reported for standard glass formers)
and $\Delta\sigma=0.1$, and discuss the solutions as functions of varying
distance parameter $\sigma$. The
non-universal material-dependent exponents are thus fixed as
$b\approx0.558$ and $\gamma\approx2.537$, and
$B\approx0.918$ \cite{Goetze90}.

\begin{figure}
\includegraphics[width=.9\linewidth]{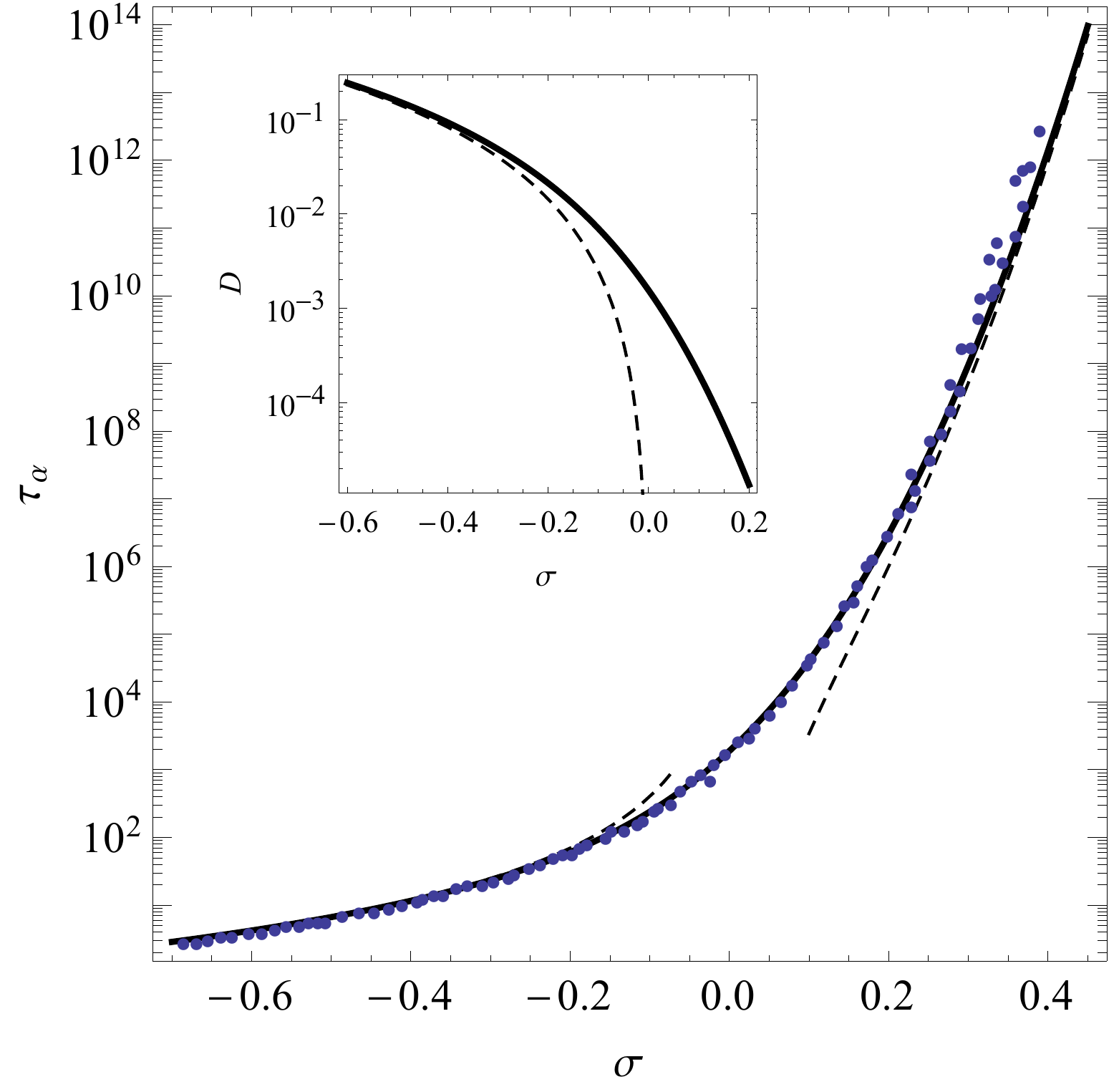}
\caption{Relaxation time $\tau_\alpha$ from the sSBR theory,
as a function of the distance $\sigma$
to the ideal-MCT glass
transition at $T_c$ ($\sigma<0$: liquid, $\sigma>0$: glass).
Inset, diffusion coefficient $D$ obtained from the theory, see text.
Symbols are dielectric-spectroscopy data for propylene carbonate from Ref.~\protect\cite{Lunkenheimer00}.
}
\label{fig:ta-vs-s}
\end{figure}

Strictly speaking, the SBR theory cannot make statements about the final structural ($\alpha$) relaxation and its time scale $\tau_\alpha$. But it is often observed that the shape of the $\alpha$ relaxation is almost invariant under a change of temperature, up to a shift of time scales. This time-temperature superposition principle emerges from MCT in the liquid \cite{Goetze09}. Assuming it to hold approximately also for $\sigma>0$, we may regard the time scale $t_\sigma'$ as a proxy for $\tau_\alpha$. Inserting the von~Schweidler asymptote into Eq.~\eqref{gsigt},
\begin{equation}\label{tau}
  \tau_\alpha\approx B_\lambda^{-1/b}\left[\int_{-\infty}^0
  \frac{ds}{\sqrt{2\pi}\,\Delta\sigma}e^{-\frac{(s-\sigma)^2}{2\Delta\sigma^2}}
  |s|^{b\gamma}\right]^{-1/b}\,.
\end{equation}
For $\sigma\to-\infty$, this recovers $\tau_\alpha\propto|\sigma|^{-\gamma}$, the well-known power-law growth of the structural relaxation time as the MCT-$T_c$ is approached from above. Around $T_c$, it crosses over to exponential growth, since for $\sigma\to\infty$,
\begin{equation}\label{tauglass}
  \tau_\alpha\propto 
%B_\lambda^{-1/b} 
e^{\frac{\sigma^2}{2b\Delta\sigma^2}}
  \sigma^{\gamma+1/b} \,.
%\times\\ \times\left[
%  \Gamma(1+b\gamma)^{-1/b}(2\pi)^{1/2b}\Delta\sigma^{-2\gamma-2/b}\right]\,.
\end{equation}
Figure~\ref{fig:ta-vs-s} shows $\tau_\alpha$ as a function of distance parameter $\sigma$, evaluated from the solution of Eq.~\eqref{tau}. The two asymptotes are shown as dashed lines. They reveal the difficulty in finding the correct asymptotic laws of ideal MCT in experiment: for large $|\sigma|$, pre-asymptotic correction effects will dominate, but for sufficiently small $|\sigma|$, the behavior is dominated by the crossover between the asymptotic laws shown in the figure. For comparison, we include in Fig.~\ref{fig:ta-vs-s} exemplary experimental data, obtained by Schneider \textit{et~al.}\ \cite{Schneider99} from dielectric spectroscopy on propylene carbonate (from
Ref.~\cite{Lunkenheimer00}, time scaled by $t_0\approx1.7\times10^{11}\,\text{s}$, assuming $\sigma=1.47 \times (\beta-\beta_c)/\beta_c$ and $T_c=1/\beta_c=200K$ (the value $T_c=180K$ \cite{Goetze00}  would also work, albeit in a smaller region around sigma=0).

\begin{figure}
\includegraphics[width=.9\linewidth]{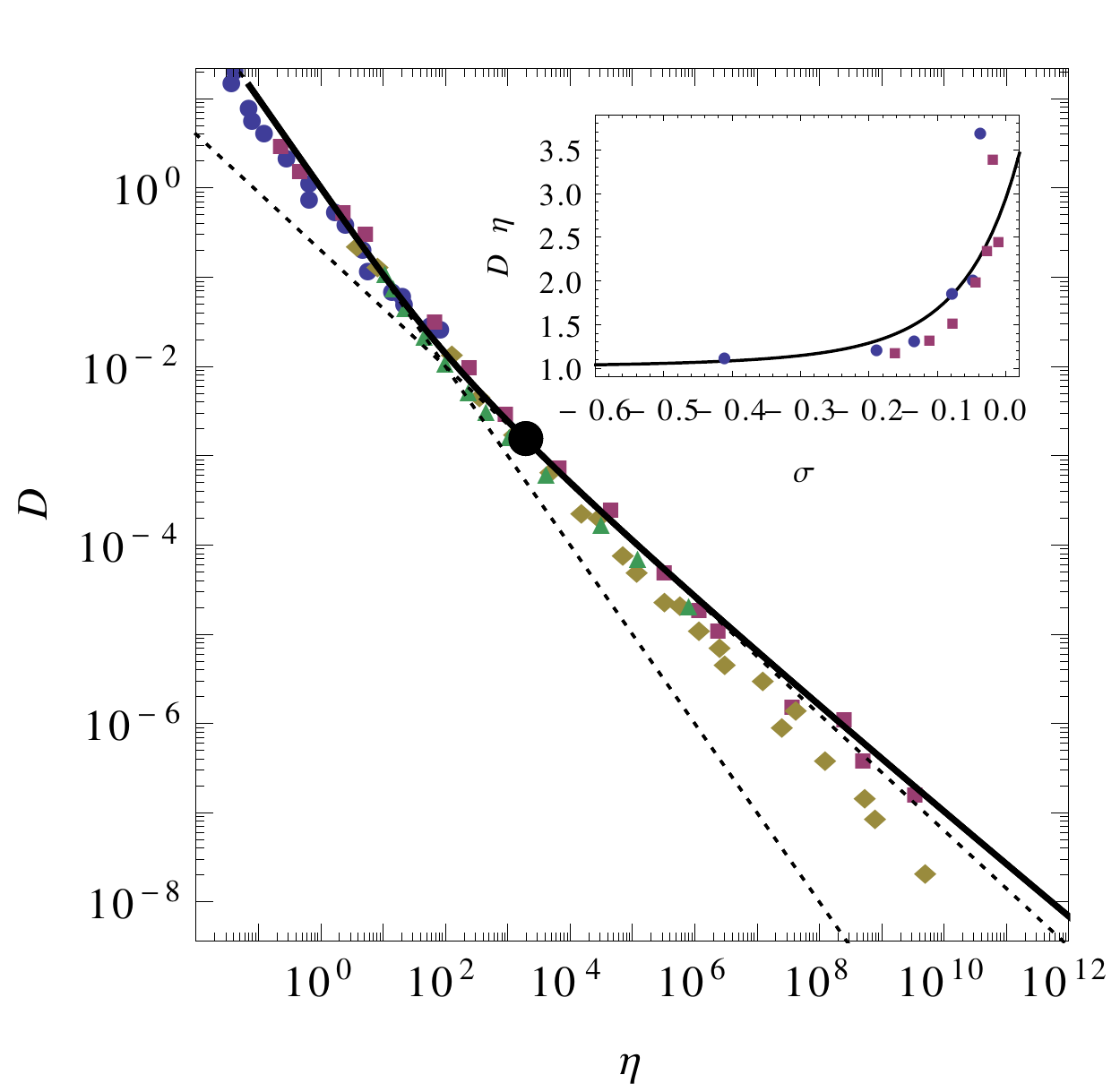}
\caption{Diffusion coefficient $D$ versus
viscosity $\eta$ in the sSBR theory (solid line).
Dashed lines indicate $D\propto\eta^{-1}$ (Stokes-Einstein
relation), and a fit with $D\propto\eta^{-0.65}$ (fractional SE relation).
A large circle marks $T=T_c$.
Smaller symbols: experimental data for o-terphenyl from Ref.~\cite{Lohfink92} (circles: tracer diffusion at $T\gtrsim T_c$; squares: diffusion of flourescent ACR dye; diamonds: TTI dye); simulation results for a harmonic-sphere mixture (triangles, Ref.~\cite{Flenner13}).
Inset: $D\cdot\tau$ for $\sigma<0$ with simulation data (circles: binary Lennard-Jones mixture \protect\cite{Kob94}; squares: Brownian hard spheres \protect\cite{Voigtmann04}).
}
\label{fig:ser}
\end{figure}

To address the Stokes-Einstein relation, let us take proxies for the viscosity $\eta$ and the diffusion coefficient $D$, by setting $\eta\sim\tau_\alpha$ and $D\sim[1/\tau_{\alpha,\sigma+s}]$. Although one needs to be careful about such reasoning in detail \cite{Koehler13}, this captures qualitatively the notion that viscosity is connected to the slow relaxation time, while diffusion is proportional to a local rate.
Explicitly,
\begin{equation}\label{D}
  D\sim B_\lambda^{1/b}\int_{-\infty}^0\frac{ds}{\sqrt{2\pi}\,\Delta\sigma}
  e^{-\frac{(s-\sigma)^2}{2\Delta\sigma^2}}|s|^\gamma\,.
\end{equation}
Quite naturally, one then obtains a crossover describing the violation of the SE relation below $T_c$. Far above $T_c$, the distribution of the $\tau_{\alpha,\sigma+s}$ is dominated by the mean, and $D\cdot\eta\sim\text{const.}$ results. For $\sigma\gg0$, we get from Eq.~\eqref{D}
\begin{equation}\label{Dglass}
  D \propto 
%B_\lambda^{1/b}
e^{\frac{-\sigma^2}{2\Delta\sigma^2}}
  \sigma^{-\gamma-1}
%\left[\Gamma(1+\gamma)(2\pi)^{1/2}\Delta\sigma^{1+2\gamma}\right]
\,.
\end{equation}
From Eqs.~\eqref{tauglass} and \eqref{Dglass}, both $1/D$ and $\eta$ grow exponentially for $T<T_c$, but with different exponents. This leads to a fractional SE relation, with logarithmic correction,
\begin{equation}\label{se}
  D\cdot\eta^b \propto (\ln\eta)^{\frac{b\gamma-1}{2}}
  \,.
 % \left(\frac{2b}{\Delta\sigma^2}\right)^{\frac{b\gamma-1}{2}}
 % \frac{\Gamma(1+\gamma)}{\Gamma(1+b\gamma)}B_\lambda^{\frac{1-b}{b}}\,.
\end{equation}
The behavior of $D$ as a function of $\eta$ is shown in Fig.~\ref{fig:ser}. The solution according to the sSBR clearly exhibits the two asymptotes discussed here. The original SE relation only holds for a rather limited regime of the slow dynamics, and already at $T_c$, noticeable deviations are seen. This agrees well with simulation results. For large viscosities, i.e., below $T_c$, a fractional SE relation, $D\approx\eta^{-x}$ with an effective exponent $x\approx0.65$ can be fitted over about $6$ orders of magnitude in $\eta$. Note that $x$ differs from the exponent $b\approx0.558$ appearing in Eq.~\eqref{se}, owing to strong logarithmic corrections. In experiments on OTP using two chemically different tracer molecules \cite{Lohfink92}, shown as symbols in Fig.~\ref{fig:ser}, a value $x\approx0.79$ was reported. But recall that the exponents are material-dependent. Also, far below $T_c$ the diffusivities of different tracers can strongly decouple \cite{Bartsch10}. This is beyond the scope of the asymptotic theory we discuss here.

Equilibrated computer simulation data usually access the regime $T\gtrsim T_c$ (or $\varphi\lesssim\varphi_c$ for the packing fraction of hard-core particles). There, an increase in $D\cdot\tau_\alpha$ of about a factor $3$ is often reported. This agrees quantitatively with our results, as shown in the inset of Fig.~\ref{fig:ser}. There, exemplary simulation data have been added for the Kob/Andersen binary Lennard-Jones mixture \cite{Kob94} (assuming $\sigma\approx0.5(T_c-T)/T_c$) and for Brownian-dynamics simulation of quasi-hard spheres \cite{Voigtmann04} (assuming $\sigma\approx(\varphi-\varphi_c)/\varphi_c$). Also, recent simulations of a harmonic-sphere mixture \cite{Flenner13} (triangles in Fig.~\ref{fig:ser}) indicate a fractional SE relation. There it was emphasized that the breakdown of the SE relation sets in significantly above $T_c$. Our sSBR calculations support this.

A previous extended MCT by Chong \cite{Chong08} finds a fractional SE relation only as a crossover to another regime with $x=1$ \cite{Chong09}, by assuming single-particle hopping processes. But the absence of a tracer-mass dependence (isotope effect) for diffusion in metallic glass formers indicates that diffusion below $T_c$ continues to be highly collective \cite{Ehmler98}. This is naturally expressed in the SBR theory.

\begin{figure}
\includegraphics[width=.9\linewidth]{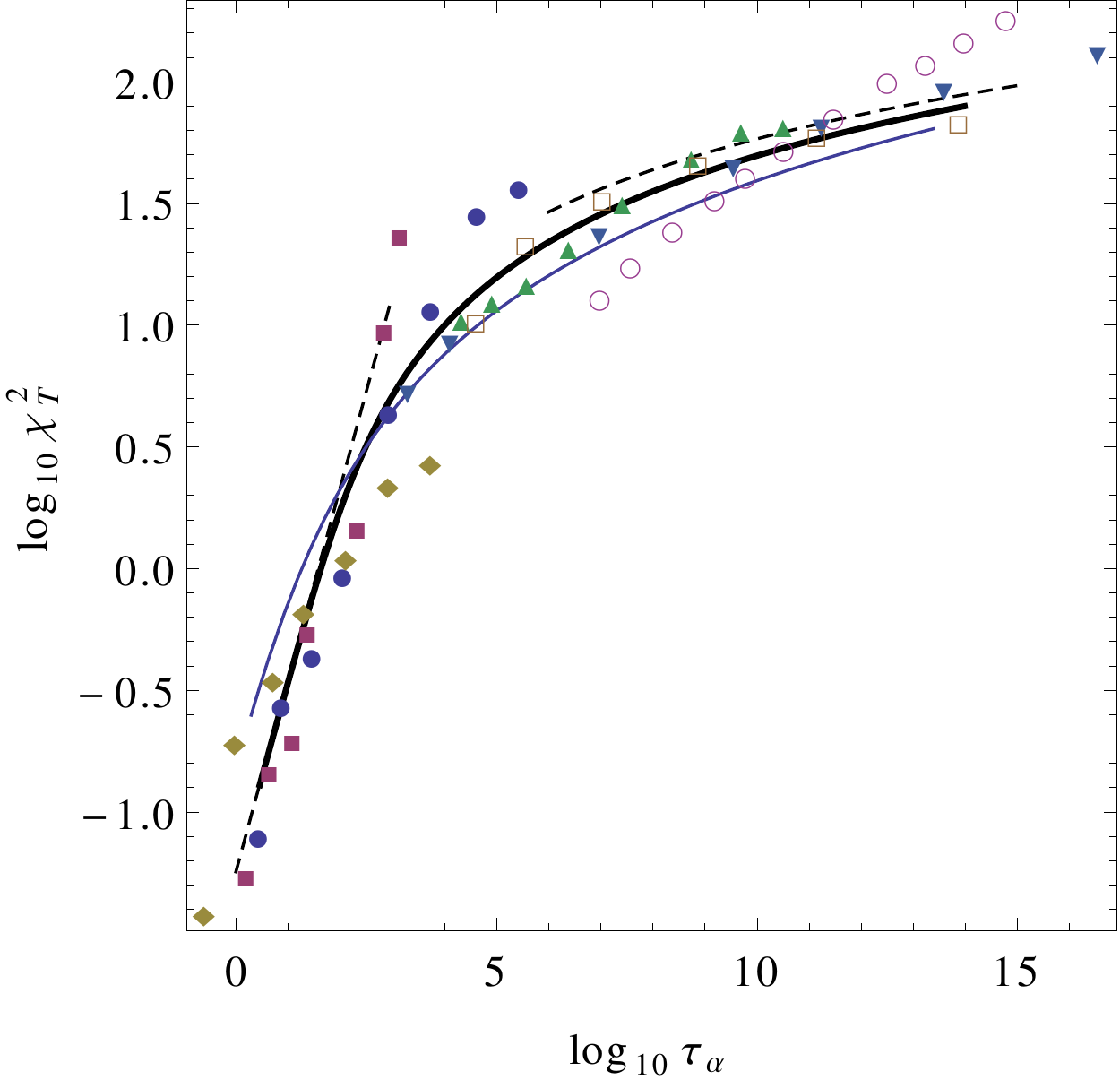}
\caption{Peak of the squared thermal susceptibility $\chi_T^2$ as a function
of relaxation time $\tau_\alpha$ from sSBR theory (solid line); asymptotes as dashed lines. Symbols: experimental data extracted from Ref.~\protect\cite{DalleFerrier07}, shifted by arbitrary factors along the vertical (circles: Lennard Jones mixture; squares: hard spheres; BKS silica: diamonds; triangles: propylene carbonate; inverted triangles: glycerol; open circles: OTP; open squares: salol). Thin line: expression proposed in Ref.~\protect\cite{DalleFerrier07}.}
\label{fig:XT2-vs-ta}
\end{figure}

We now turn to a discussion of the thermal susceptibility studied extensively in recent years \cite{Berthier05}. It is defined as the temperature derivative of the density correlation function,
$
  \chi_T(t)=(d/dT)\Phi(t)
$.
This quantity typically displays a peak as a function of time in the structural-relaxation regime, whose increasing height has been associated to the increase in the size of dynamical heterogeneities. Observing approximate time-temperature superposition, $\Phi(t)\approx C(t/\tau_\alpha)$, the dependence of $\chi_T(t)$ on $T$ is given through the dependence of $\tau_\alpha$ on $\sigma$. The maximum of $\chi_T(t)$ is then given at constant $t/\tau_\alpha$, so that the $\sigma$-dependence of the squared height is just
\begin{equation}
  \chi_T^2\propto\left(\frac1{\tau_\alpha}\frac{d\tau_\alpha}{d\sigma}\right)^2 \, ,
\end{equation}
and the sSBR prediction can be obtained from Eq.~\eqref{tau}.
We get for $\sigma\ll0$ the known ideal-MCT power law \cite{Berthier07}
$
  \chi_T^2\sim\gamma^2 B_\lambda^{2/b\gamma}\tau_\alpha^{2/\gamma}
  %\,,\qquad\tau_\alpha\ll1
$,
and for the glass state at $T<T_c$,
\begin{equation}
  \chi_T^2\sim\frac2{b\Delta\sigma^2}\ln\tau_\alpha
  -\frac{1+b\gamma}{b^2\Delta\sigma^2}\ln\ln\tau_\alpha\,,
  \qquad\tau_\alpha\gg1\,.
\end{equation}
Figure~\ref{fig:XT2-vs-ta} shows the sSBR result (full line) together with the liquid and glass asymptotes (dashed). Again the ideal-MCT asymptote already breaks down while $\tau_\alpha$ is an order of magnitude lower than the value at $T_c$; there is a large crossover window around $T_c$.

The crossover of the susceptibility from initial fast
to much slower logarithmic growth around $T_c$ was found
empirically before.
Dalle-Ferrier \textit{et~al.}\ \cite{DalleFerrier07} suggested
$\tau_\alpha\approx A(N_\text{corr}/N_0)^\gamma\,\exp[(N_\text{corr}/N_0)^\psi]$, where $N_\text{corr}\sim|\partial\ln\tau_\alpha/\partial\ln T|\sim T\chi_T$ is the number of correlated atoms that characterize the dynamics, and $\psi\approx1.4$ was found to fit experimental data well.
Assuming $\chi_T^2\propto N_\text{corr}^2$ with some arbitrary prefactor, we have added these data to Fig.~\ref{fig:XT2-vs-ta} (extracted from Ref.~\cite{DalleFerrier07}; see there for original references). For comparison, the empirical relation proposed by Dalle-Ferrier \textit{et~al.}\ is also shown (thin line). It qualitatively follows the result of the sSBR theory, and adjusting its parameters freely, one can make the two curves agree over the range of $\tau_\alpha$ shown.

In conclusion, we have shown that the application of field-theoretical concepts in the context of MCT's asymptotic laws provides a qualitative explanation for a number of generic features found in glass-forming liquids around the MCT-$T_c$. This includes the crossover in the growth of $\tau_\alpha$, the emergence of a fractional Stokes-Einstein relation, and the two asymptotic regimes for the growth of the thermal susceptibility that is commonly associated with dynamic heterogeneities.

The SBR theory is asymptotic in nature, i.e., it is expected to work in some region around $T_c$ and not necessarily deep inside the glass. It describes the features of the intermediate-time $\beta$ relaxation and thus the initial relaxation from the nonergodicity plateau. In the discussion, we have assumed that this provides a qualitative proxy for the structural relaxation time $\tau_\alpha$. Judging by the comparison to experimental data this appears reasonable, although the emergence of secondary relaxation wings that are prominent in dielectric spectra \cite{Lunkenheimer00} could in principle destroy such a connection.

In discussing the SE violation it should be noted that the experimental facts may not be clear -- for some metallic glass formers, the SE relation was reported to be well fulfilled for one atomic species even at $T\ll T_c$ \cite{Bartsch10}, but the precise determination of $D$ and $\eta$ is difficult. Still, the SBR theory provides a robust and intuitive explanation, why a breakdown of the SE relation can set in for $T\approx T_c$. It predicts the exponent of the fractional SE relation to be intimately connected to the MCT dynamics through the von~Schweidler law, although logarithmic corrections hide this connection.

\begin{acknowledgments}
We thank L.~Leuzzi, M.~Fuchs and A.~Meyer for discussions.
\end{acknowledgments}

\bibliographystyle{apsrev4-1}
\bibliography{lit}

\end{document}